\documentclass[reprint,twocolumn,superscriptaddress,secnumarabic,amssymb,nobibnotes,aps,prr]{revtex4-2}
\usepackage{graphicx}% Include figure files
\usepackage{dcolumn}% Align table columns on decimal point
\usepackage{bm}% bold math
\usepackage{xcolor}
\usepackage[export]{adjustbox}
\usepackage{float}
\usepackage{tabularx}
\usepackage{multirow}
\usepackage{amsmath}
\usepackage{braket}
\usepackage{array}
\usepackage{ulem}
\usepackage{comment}

\newcolumntype{C}[1]{>{\centering\let\newline\\\arraybackslash\hspace{0pt}}m{#1}}

\begin{document}

%\preprint{}

\title{High-resolution valence band RIXS at the actinide \textit{M}$_{4,5}$-edges}

\author{Martin Sundermann}
\affiliation{Max Planck Institute for Chemical Physics of Solids, N{\"o}thnitzer Stra{\ss}e 40, 01187 Dresden, Germany}
\affiliation{PETRA III, Deutsches Elektronen-Synchrotron DESY, Notkestra{\ss}e 85, 22607 Hamburg, Germany}

\author{Henrik Hahn}
\affiliation{Institute for Theoretical Physics, Heidelberg University, Philosophenweg 19, 69120 Heidelberg, Germany}

\author{Denise S. Christovam}
\altaffiliation{present address: Max Planck Institute for Solid State Research, Heisenbergstra{\ss}e 1, 70569 Stuttgart, Germany}
\affiliation{Max Planck Institute for Chemical Physics of Solids, N{\"o}thnitzer Stra{\ss}e 40, 01187 Dresden, Germany}

\author{Maurits~W.~Haverkort}
\affiliation{Institute for Theoretical Physics, Heidelberg University, Philosophenweg 19, 69120 Heidelberg, Germany}

\author{Roberto Caciuffo}
\affiliation{European Commission, Joint Research Centre, Postfach 2340, DE-76125 Karlsruhe, Germany}
\affiliation{Dipartimento di Scienze Matematiche, Fisiche e Informatiche, Universit\`{a} di Parma, Parco Area Delle Scienze 7/A, 43124 Parma, Italy}

\author{Bernhard~Keimer}
\affiliation{Max Planck Institute for Solid State Research, Heisenbergstra{\ss}e 1, 70569 Stuttgart, Germany}

\author{Liu~Hao~Tjeng}
\affiliation{Max Planck Institute for Chemical Physics of Solids, N{\"o}thnitzer Stra{\ss}e 40, 01187 Dresden, Germany}

\author{Andrea~Severing}
\altaffiliation{corresponding author: andrea.severing@cpfs.mpg.de}
\affiliation{Max Planck Institute for Chemical Physics of Solids, N{\"o}thnitzer Stra{\ss}e 40, 01187 Dresden, Germany}
\affiliation{Institute of Physics II, University of Cologne, Z\"{u}lpicher Stra{\ss}e 77, 50937 Cologne, Germany}

\author{Hlynur~Gretarsson}
\altaffiliation{corresponding author: hlynur.gretarsson@desy.de}
\affiliation{PETRA III, Deutsches Elektronen-Synchrotron DESY, Notkestra{\ss}e 85, 22607 Hamburg, Germany}
\affiliation{Max Planck Institute for Solid State Research, Heisenbergstra{\ss}e 1, 70569 Stuttgart, Germany}
%\email{} 
\date{\today}

\begin{abstract}
Understanding the electronic structure of actinide materials is crucial for both fundamental research and nuclear applications. The partially filled 5\textit{f} shells exhibit complex behavior due to strong correlations and ligand hybridization, requiring advanced spectroscopic techniques. Here, we report on the development and application of high-resolution valence-band resonant inelastic x-ray spectroscopy (VB-RIXS) experiments at the uranium \textit{M}$_{4,5}$ edges (3551 and 3725\,eV). We present data of UO$_2$, a well-established model actinide compound. VB-RIXS is particularly well suited for probing the 5\textit{f}-shell electronic structure, as it probes, in contrast to core-to-core RIXS,  excitations without leaving a high-energy core hole in the final state. In VB-RIXS, we achieve energy resolutions of 50\,meV (\textit{M}$_5$) and 90\,meV (\textit{M}$_4$), enabling the resolution of multiplet excitations and crystal-field effects, as well as charge-transfer and fluorescence-like features with unprecedented clarity. As such, high resolution VB-RIXS offers direct insights into both low-energy, near ground-state properties and high-energy hybridization and covalency effects. Our results demonstrate the power of VB-RIXS as a versatile and powerful tool for probing the strongly correlated electronic structure of actinide materials, providing essential input for quantitative modeling and the validation of theoretical concepts.

\end{abstract} 

\maketitle

%%%%%%%%%%%%%%%%%%%%%%%%%%%%%%%%%%%%%%%%%%%%%%%%%%%%%%%%%%%%%
\section{Introduction}
Actinide chemistry and physics play a crucial role in nuclear energy, radioactive waste management, and fundamental science. The actinides, including uranium, plutonium, and thorium, are key elements in nuclear fuel cycles, and understanding their properties is essential for improving reactor efficiency, developing safer fuels, and managing long-lived radioactive waste. From a scientific perspective, the partially filled 5$f$ shells in actinides exhibit complex electronic structures due to strong electron correlation, making actinide materials very challenging but also valuable study objects for advancing our modeling techniques in the field of solid state physics and solid state chemistry.

The 5\textit{f} shell is spatially more extended than the 4\textit{f} shell of the rare earths, yet more localized than the \textit{d} shells of the transition metals, so that the clear hierarchy of interactions, Coulomb, spin orbit and crystal-field, is no longer well defined. Additionally, the greater spatial extent of the 5\textit{f} shell leads to substantially stronger hybridization with its surrounding compared to the  4\textit{f} shell in rare earth compounds. 

\begin{figure*}[t]
	\begin{center}
		\includegraphics[width=1.99\columnwidth]{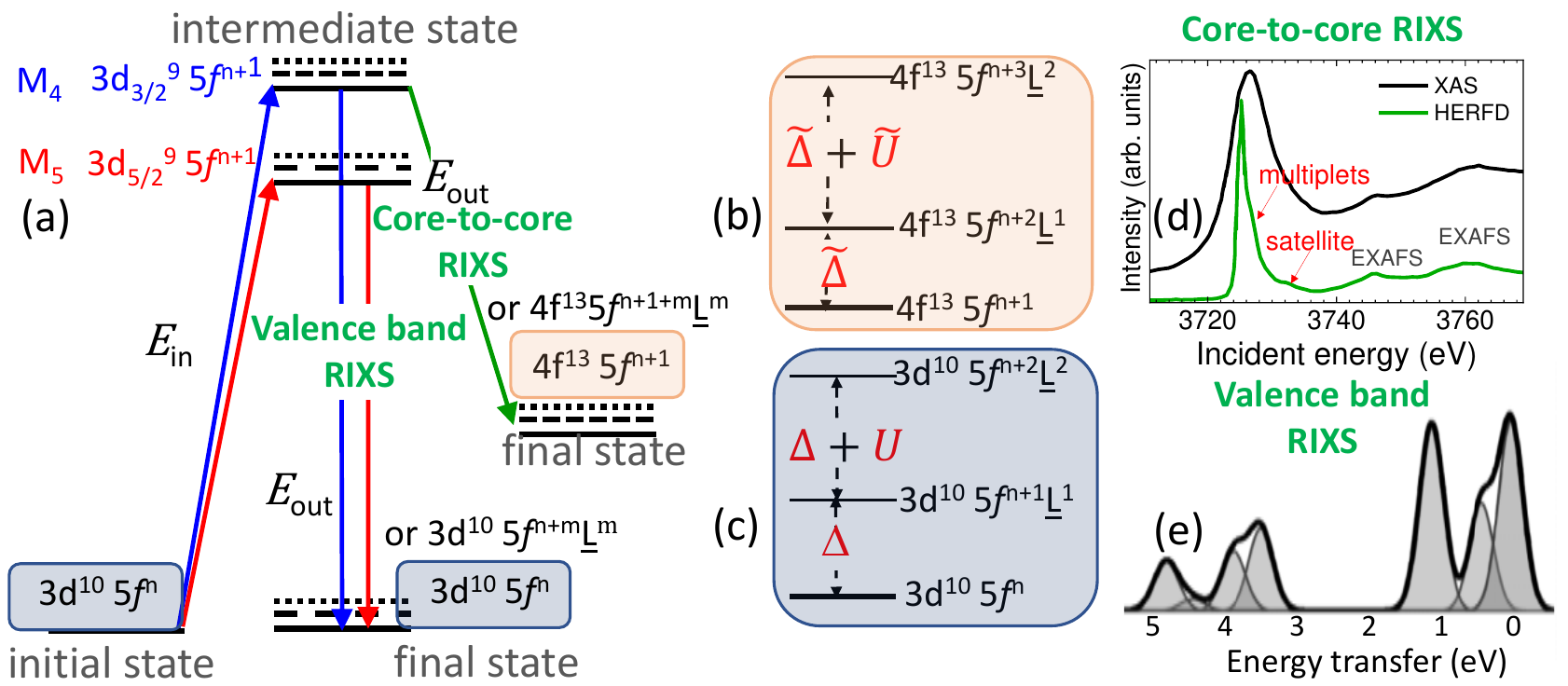}
	\end{center}
	\caption{(a) Scattering channels in core-to-core RIXS and VB-RIXS. (b) Energy-level diagram of VB-RIXS final state. (c) Energy-level diagrams of the VB-RIXS and core-to-core RIXS initial states as well as core-to-core RIXS final state.  (d) M4-edge XAS (black line) compared to HERFD spectra in M4-edge core-to-core RIXS (green line). (e) Cartoonlike VB-RIXS spectrum with multiplet and crystal-field excitations.}  
	\label{method}
\end{figure*}

To develop a many-body model that properly describes the electronic structure, it is essential to determine the oxidation state or, equivalently, the formal 5$f^n$ configuration that defines the quantum numbers. Equally important is identifying the symmetry of the crystal-field-split Hund's rule ground state within the specific crystalline environment or coordination geometry. Additionally, the degree of covalency, arising from hopping between the 5\textit{f} electrons and the surrounding ligands as well as between the 5\textit{f} electrons at different sites, plays a crucial role. A comprehensive understanding of these factors is fundamental for constructing a sound electronic structure model, which can then be used to address key physical or chemical questions about the material under investigation. 

Here we will address the example of UO$_2$, which is a well studied material due to its important role as the standard nuclear fuel in commercial nuclear reactors. It also exhibits interesting electronic and magnetic properties, including Mott-insulating behavior with an excitation gap of about 2\,eV\,\cite{Schoenes1980,Kudin2002} and triple-$\bf{k}$ antiferromagnetic ordering below $T_\text{N}$\,=\,30.8\,K\,\cite{Blackburn2005,Santini2009}, accompanied by a Jahn-Teller distortion of the oxygen and the ordering of uranium electric quadrupoles\,\cite{Wilkins2004,Wilkins2006}. 

In the following, we explain why commonly used techniques, such as photoelectron and x-ray absorption spectroscopy (PES/XAS), are not sufficient for understanding uranium compounds (and early actinide materials in general), and why it is necessary to employ more advanced methods. 

From the perspective of electronic structure, uranium in UO$_2$ presents a particularly challenging case. As an early actinide, it contains a large number of holes in the 5\textit{f} shell, resulting in a large 5\textit{f}-level degeneracy within a many-body framework. This leads to large effective inter-configuration transfer integrals\,\cite{Gunnarsson1983,Yamazaki1991} and necessitates the consideration of multiple 5\textit{f}$^{n+m}$\textit{\underline{L}}$^m$ configurations, with \underline{\textit{L}}$^m$ denoting $m$ holes on the ligand site and  ($m$ = 0, 1, 2, 3, ...), especially when interpreting photoemission spectra\,\cite{Kotani1988,Yamazaki1991,Kotani1992} where the added Coulomb interactions due to the photoemission hole significantly modify the relative energies of the electronic configurations\,\cite{Gunnarsson1983}. This reordering of energy levels, combined with the inherent degeneracy, complicates spectral simulations and hinders the reliable extraction of key model parameters, such as the degree of covalency. A similar situation is encountered in VO$_2$, where the large number of $d$ shell holes in vanadium necessitates the inclusion of five to six configurations to accurately reproduce the vanadium core-level spectra \cite{Yamaguchi2024}. These complexities may even call into question the validity of the commonly employed single-site approximation in such theoretical treatments.

For early actinide compounds, and similarly for early transition metal compounds, charge-conserving probes may be the more suitable choice. In x-ray absorption spectroscopy (XAS), where a core electron is excited into the 5\textit{f} shell, the energy structure of the configurations in the final state closely resembles that of the initial state\,\cite{Gunnarsson1983, Groot1994}. As a result, the number of configurations required in simulations can be kept relatively small. This relies on the assumption that the Coulomb interaction between the core hole and the 5\textit{f} electron is of comparable magnitude but opposite in sign to that between two 5\textit{f} electrons, such that the absorption process does not introduce significant additional Coulomb interactions. Under these conditions, the characteristic multiplet structures observed in XAS can be even sensitive to the symmetry of the ground state\,\cite{Groot1994}. However, in practice, the XAS spectra of uranium compounds are often very broad and featureless\,\cite{Groot2008}, providing limited insight into key quantities such as covalency, formal configuration, or the symmetry of the ground state. While sum-rule and branching-ratio analyses offer alternative routes \cite{Laan1996,Wilhelm2018,Wilhelm2023}, their precision is limited in non-ionic systems exhibiting strong intermediate- or mixed-valent character.

A more promising charge neutral spectroscopic probe is offered by resonant inelastic x-ray scattering (RIXS) techniques. RIXS is a more direct probe to multiplet structures and satellites related to covalency. In the following, we discuss the differences between well-established core-to-core RIXS and valence-band (VB) RIXS, the latter representing a novel approach in the field of uranium and actinide research. We  demonstrate the power of \textit{M}$_{4,5}$-edge VB-RIXS using UO$_2$ as a model system.

\section{Methods: core-to-core RIXS versus VB-RIXS}
The most significant difference between core-to-core RIXS and VB-RIXS lies in their final states. In core-to-core RIXS, the final states contains a core-hole and does not involve the same set of 5\textit{f} configurations as the initial (ground) state. In contrast, VB-RIXS results in final states without a core hole, allowing its spectral features to be directly related to the initial state configurations.  

Over the past 10 to 15 years, significant advances have been made in core-to-core RIXS techniques and spectral modeling, aimed at studying metal-ligand and actinide bonding, atomic arrangements, and the electronic structure, including correlation effects, in actinide and uranium complexes, see Ref.s\,\cite{Kvashnina2013,Kvashnina2014,Sergentu2018,Polly2021,Amidani2021,Kvashnina2022,Misael2023,Silva2024,Bagus2024,Burrow2024,Schacherl2025} and references therein. Figure\,\ref{method}\,(a) illustrates the core-to-core RIXS process: Following the XAS excitation from the 3\textit{d} to the 5\textit{f} shell (3\textit{d}$^{10}$5\textit{f}$^n$\,$\rightarrow$\,3\textit{d}$^9$5\textit{f}$^{n+1}$), the subsequent decay process leads into a set of excited states 3\textit{d}$^{10}$4\textit{f}$^{13}$5\textit{f}$^{n+1}$, approximately 385\,eV above the ground state and featuring a core hole in the 4\textit{f} shell, see Fig.\,\ref{method}\,(b). Here the 4\textit{f} core hole determines the lifetime broadening so that the originally broad ($\approx$4\,eV) XAS line is significantly narrowed below the core-hole life time broadening of the 3$d_{3/2}$ or 3$d_{5/2}$ core-holes, yielding high-resolution XAS-like spectra. 
%>>>>>>>>>>>>>>>>>>>>>>>>>

Using a clever cut in the energy transfer--incident energy map, namely the cut for constant final energy, reduces the broadening further,\,\cite{Kvashnina2022} and references therein. This technique, known as high energy resolution fluorescence detection (HERFD), allows for the detection of shifts of 1 to 2\,eV in the white line position due to different 5\textit{f}$^n$ configurations, as well as smaller shifts ($\approx$0.1\,eV) due to coordination geometry and ligand-field splitting. Additionally,  HERFD spectra reveal shoulders arising from multiplet and crystal-field effects, as well as satellites due to hopping or hybridization with surrounding ligands (charge transfer). The corresponding final state for the charge transfer satellites is  4\textit{f}$^{13}$5\textit{f}$^{n+m}$\underline{\textit{L}}$^m$ with \underline{\textit{L}}$^m$ denoting $m$ holes on the ligand site. Figure\,\ref{method}\,(d) compares \textit{M}$_4$-edge XAS and HERFD spectra of UO$_2$. The method is also applicable to transuranium compounds at specially dedicated beamlines\,\cite{Schacherl2025}.

While HERFD reveals the presence of multiplets and satellites, the final state nevertheless still contains a core-hole. Furthermore, multiplet structures, let alone crystal-field splittings are not resolved, due to the lifetime broadening of the core-hole. The satellites are also weak and lack substructures so that the data impose limited constraints on the simulations. To overcome these limitations, we turn to valence band resonant inelastic x-ray scattering (VB-RIXS) at the U\,\textit{M}$_{4,5}$-edges. 

%\subsection{VB-RIXS}

The scattering process of VB-RIXS is also depicted in Fig.\,\ref{method}\,(a). As in core-to-core RIXS, an electron is excited from 3\textit{d} to the 5\textit{f} shell but then de-excites into the ground state 3\textit{d}$^{10}$5\textit{f}$^n$ (elastic scattering) or a near ground state excitations of the same configuration  (crystal-field split multiplet scattering), or charge excitations, e.g., into another 5\textit{f} configuration 3\textit{d}$^{10}$5\textit{f}$^{n+m}$\textit{\underline{L}$^m$} due to hybridization (charge transfer), see Fig.\,\ref{method}\,(c). For a fixed incident energy, \textit{E}$_{in}$, that can be tuned to the resonances of the \textit{M}$_5$ or \textit{M}$_4$ absorption edges, \textit{E}$_\text{res}$, or to their vicinities, a spectrum is recorded as a  function of energy transfer. Figure\,\ref{method}\,(e), shows such a \textit{model} VB-RIXS spectrum that, dependent on the resolution and energy range, is reminiscent of inelastic neutron scattering data.

The energy-level diagrams, featuring for simplicity only three 5\textit{f} configurations, in Fig.\,\ref{method}\,(b) and (c) illustrate the more direct access to charge-transfer processes in VB-RIXS compared to core-to-core RIXS. In both cases, the initial state is a hybridized state composed of 5\textit{f}$^n$, 5\textit{f}$^{n+1}$\underline{\textit{L}}$^1$, and 5\textit{f}$^{n+2}$\underline{\textit{L}}$^2$ configurations, with energy differences determined by the charge-transfer energies $\Delta$ and $\Delta$\,+\,\textit{U}, where \textit{U} denotes the \textit{5f-5f} Coulomb interaction and \underline{\textit{L}}$^1$ and \underline{\textit{L}}$^2$ one or two ligand holes, respectively (see Fig.\,\ref{method}\,(c))\,\cite{Zaanen1985}. In core-to-core RIXS, the final state has a different 5\textit{f} configuration than the initial state and contains a core hole so that any parameters extracted from spectral features are reflecting the electronic structure in the presence of a core-hole. As a result, going back to our schematic model in Fig.\,\ref{method} the relevant final state energy parameters in core-to-core RIXS are $\tilde{\Delta}$ and $\tilde{\textit{U}}$, see Fig.\,\ref{method}\,(b), in contrast to VB-RIXS, where the final state is governed by the same $\Delta$ and \textit{U} as the initial state, see Fig.\,\ref{method}\,(c), making the interpretation of the spectrum to be directly linked to the initial many-body state problem\,\cite{Matsubara2005}.

VB-RIXS in the soft x-ray regime with resolutions of the order of 30\,meV has worked successfully for the cuprates\,\cite{Braicovich2009,Braicovich2010,Dean2012,Dean2013}, transition metal oxides\,\cite{Ghiringhelli2009,Bisogni2016,Hariki2018,Winder2020,Hariki2020} and rare earth compounds\,\cite{Amorese2018,Amorese2019}, but for U the best absorption edge had yet to be found. In principle, the following XAS edges can be used to investigate the U\,5$f$ shell with VB-RIXS: the U\,\textit{O}$_{4,5}$-edge (5$d$\,$\rightarrow$\,5$f$) at 110\,eV in the far ultraviolet, the U\,\textit{N}$_{4,5}$-edge (4$d$\,$\rightarrow$\,5$f$) at 770\,eV in the soft x-ray regime, or the U\,\textit{M}$_{4,5}$-edges (3$d$\,$\rightarrow$\,5$f$) at 3500 and 3700\,eV in the tender x-ray range. RIXS beamlines with excellent resolving power are available in both the far ultraviolet and soft x-ray regime but U\,\textit{O}-edge VB-RIXS suffers from strong elastic tails inherent to the far ultraviolet\,\cite{Wray2015a,Wray2015,Liu2022}, while U\,\textit{N}-edge VB-RIXS has only weak intensity\,\cite{Lander2021,Bright2023}, probably due to the weak \textit{N}-edge absorption. In contrast, the U\,\textit{M}$_{4,5}$-edges are strong absorption edges and offer notable experimental advantages: the higher photon energies provide greater probing depth, often eliminating the need for \textit{in-situ} cleaving. Conversely, measurements at the \textit{O} and \textit{N} edges demand more meticulous surface preparation due to their short attenuation lengths and thus higher surface sensitivity (see Table I in Ref.\,\cite{Caciuffo2023}).

\begin{figure}[b]
	\begin{center}
		\includegraphics[width=0.9\columnwidth]{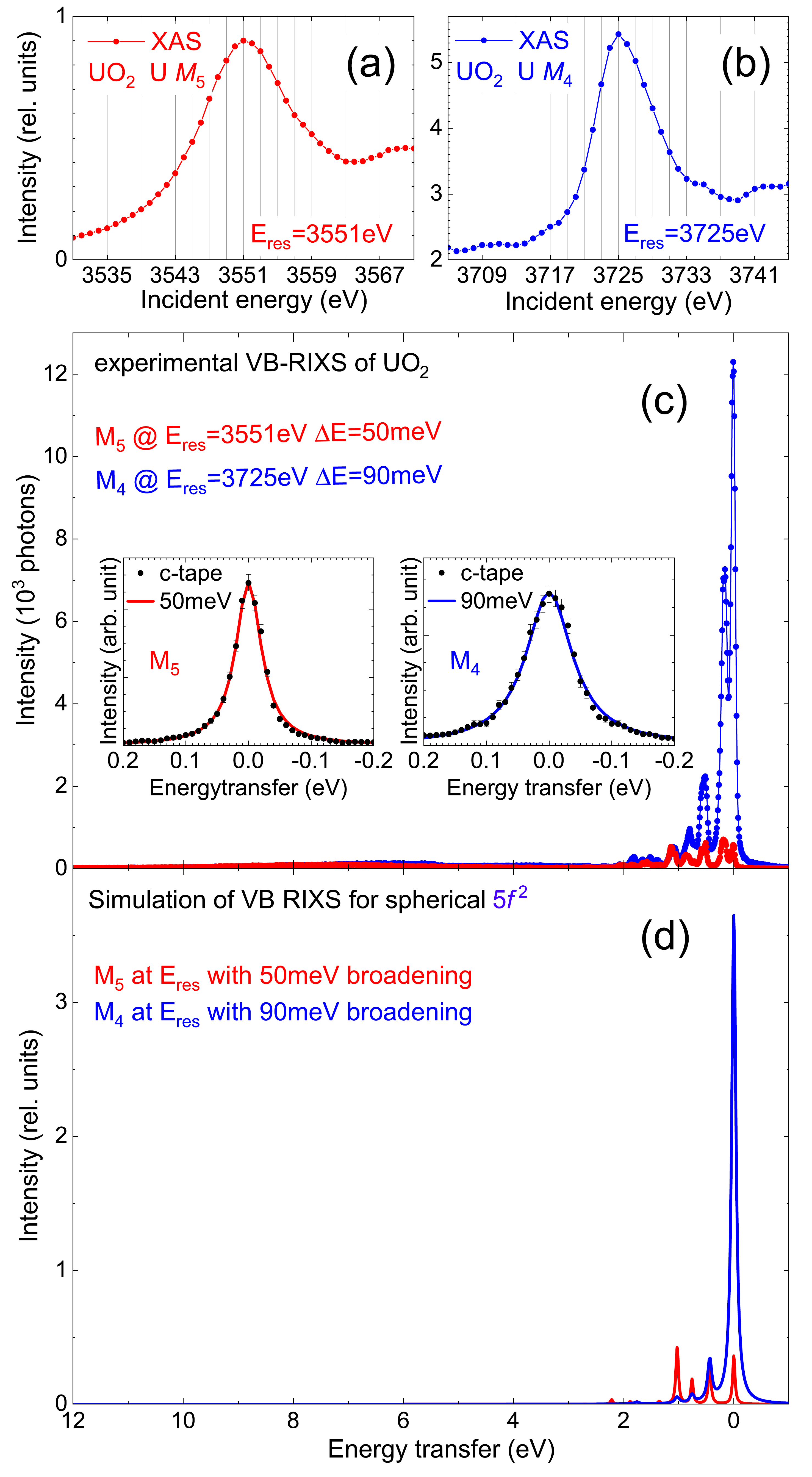}
	\end{center}
	\caption{(a), (b) XAS spectra at the U\,\textit{M}$_5$- (red dots) and \textit{M}$_4$-edges (blues dots). The vertical lines refer to incident energies used for the RIXS maps. (c) Experimental U \textit{M}$_5$- (red) and \textit{M}$_4$-edge (blue) VB-RIXS spectra measured at \textit{T}\,=\,15\,K. The intensity scales are directly comparable. Insets show elastic signals from the carbon tape at the U\,\textit{M}$_5$- (red) and \textit{M}$_4$-edges, respectively, fitted with a Lorentzian lineshape as resolution function. (d) Full multiplet VB-RIXS calculations for the 5\textit{f}$^2$2 configuration with spherical symmetry, and 50 and 90\,meV broadening, respectively.}  
	\label{RIXS_all}
\end{figure}

%%%%%%%%%%%%%%%%%%%%%%%%%%%%%%%%%%%%%%%%%%%%%%%%%%%%%%%%%%%%%%%%
\section{Experiment and Analysis}

Achieving high resolution in the tender x-ray regime is challenging because the wave-lengths are too short for using gratings as for soft x-rays, and too long for using the common silicon-based analyzer crystals as for hard x-rays. Although a total energy resolution of about 1\,eV was achieved already more than 10 years ago for tender x-rays\,\cite{Kvashnina2014}, further improvements have been hindered by the lack of suitable silicon Bragg reflections\,\cite{Kvashnina2018}. As a result, a \textit{high resolution} tender x-ray RIXS end station is quite exotic and the intermediate RIXS (IRIXS) end station, built by the Max Planck, at beamline P01 PETRA-III/DESY in Hamburg is presently the only one worldwide\,\cite{Gretarsson2020}. Through the use of quartz-based analyzer crystals, IRIXS can offer up to an order of magnitude better resolution compared to existing instruments with sufficient flux. Here flux is greatly improved by using a windowless, fully in vacuum beamline to avoid unnecessary absorption from air. 

IRIXS uses a 4-bounce Si-(111) high-resolution monochromator (HRM) to produce a narrow incident bandwidth and a spherical bent (R=1m) and diced SiO$_2$-(003) analyzer for the Rowland spectrometer. Figure\,\ref{RIXS_all}\,(a) and (b) show the elastic signal measured on a carbon tape, fitted with a Lorentzian lineshape of FWHM\,=\,50 and 90\,meV at the U \textit{M}$_5$ and \textit{M}$_4$ edge, respectively. More details on this set-up will be provided in a separate publication\,\cite{hg_newpaper}.

VB-RIXS experiments of UO$_2$ were performed at the U\,\textit{M}$_{5}$- and U\,\textit{M}$_{4}$-edges using incident energies \textit{E}$_\text{in}$ across the absorption edges in 2\,eV steps, and further away from \textit{E}$_{res}$ in 4\,eV steps. The vertical lines in Fig.\,\ref{RIXS_all}\,(c) and (d) indicate the incident energies used, as well as the respective resonance energies, \textit{E}$_\text{res}$, of 3551 and 3725\,eV. For energy calibration, a carbon tape was measured before any change in settings, such as incident energy or temperature. All spectra are aligned in energy using the carbon tape measurements.  

A UO$_2$ single crystal with a polished surface was transported in a vacuum suitcase to the beamline and inserted to the main chamber, where the pressure was 10$^{-8}$\,mbar. The [111] surface was exposed to the beam. The experiments were conducted with a scattering angle of 2$\theta$$_{in}$\,=\,90$^{\circ}$ and a specular geometry of $\phi$$_\text{sample}$\,=\,45$^{\circ}$, ensuring the momentum transfer $\vec{q}$ was parallel to (111). Energy-transfer vs.  incident energy maps were measured at \textit{T}\,=\,15\,K, i.e., below the magnetic ordering transition. For RIXS, data are taken preferably at low \textit{T}, because spectral features are usually sharper, either trough a larger optical gap, suppressed phonon contribution, or a better defined ground state. It also, generally, limits radiation damage in more sensitive samples. Additionally, \textit{M}$_5$-edge data at resonance were collected at several temperatures below and above the ordering transition to investigate potential temperature dependencies.

RIXS spectra are simulated with a full multiplet calculation using the Quanty code\,\cite{Haverkort2016}. The calculation starts with the atomic values for the 5\textit{f}$^2$ configuration from the Atomic Structure Code by Robert D. Cowan\,\cite{Cowan1981}. To simulate the effects of configuration interaction and covalency in the solid, the Slater integrals F$_k$(5\textit{f}5\textit{f}) are reduced to 50\% of their Hartree-Fock values, while F$_k$(3\textit{d}5\textit{f}) and G$_k$(3\textit{d}5\textit{f}) are reduced to 80\%\,\cite{Tanaka1994,Groot2008,Butorin2016,Agrestini2017}. Additionally, the 5\textit{f} spin-orbit coupling constants are adjusted, with $\zeta$$_{5\textit{f}}$ reduced to 90\% and $\zeta$$_{3\textit{d}}$ to 96\%. This strong reduction places the system to the intermediate coupling regime, where  \textit{J}\,=\,4 remains a good quantum number, but with a mixture of the orbital quantum numbers \textit{L}: 85\% \textit{L}\,=\,5, 14\% \textit{L}\,=\,4, and 1\% \textit{L}\,=\,3.  To account for the effect of the surrounding crystal field in O$_h$-symmetry, a point charge model is used with V$_{40}$\,=\,-134\,meV and V$_{60}$\,=30\,meV. The crystal-field parameters, described in  Ref.\,\cite{Magnani2005,Sundermann2018}, yield the $\Gamma$$_5$ triplet ground state. The life time broadening of the intermediate state also enters the RIXS calculation. Setting it to FWHM\,=\,4\,eV gives reasonable intensities. Finally, the calculations are broadened to account for the experimental resolution.

%%%%%%%%%%%%%%%%%%%% Data %%%%%%%%%%%%%%%%%%%%%%%%%%%%%%%%%%%%%%%%%%%%%%%%%%%%%%%%%%%%

\begin{figure}[t]
	\begin{center}
		\includegraphics[width=0.9\columnwidth]{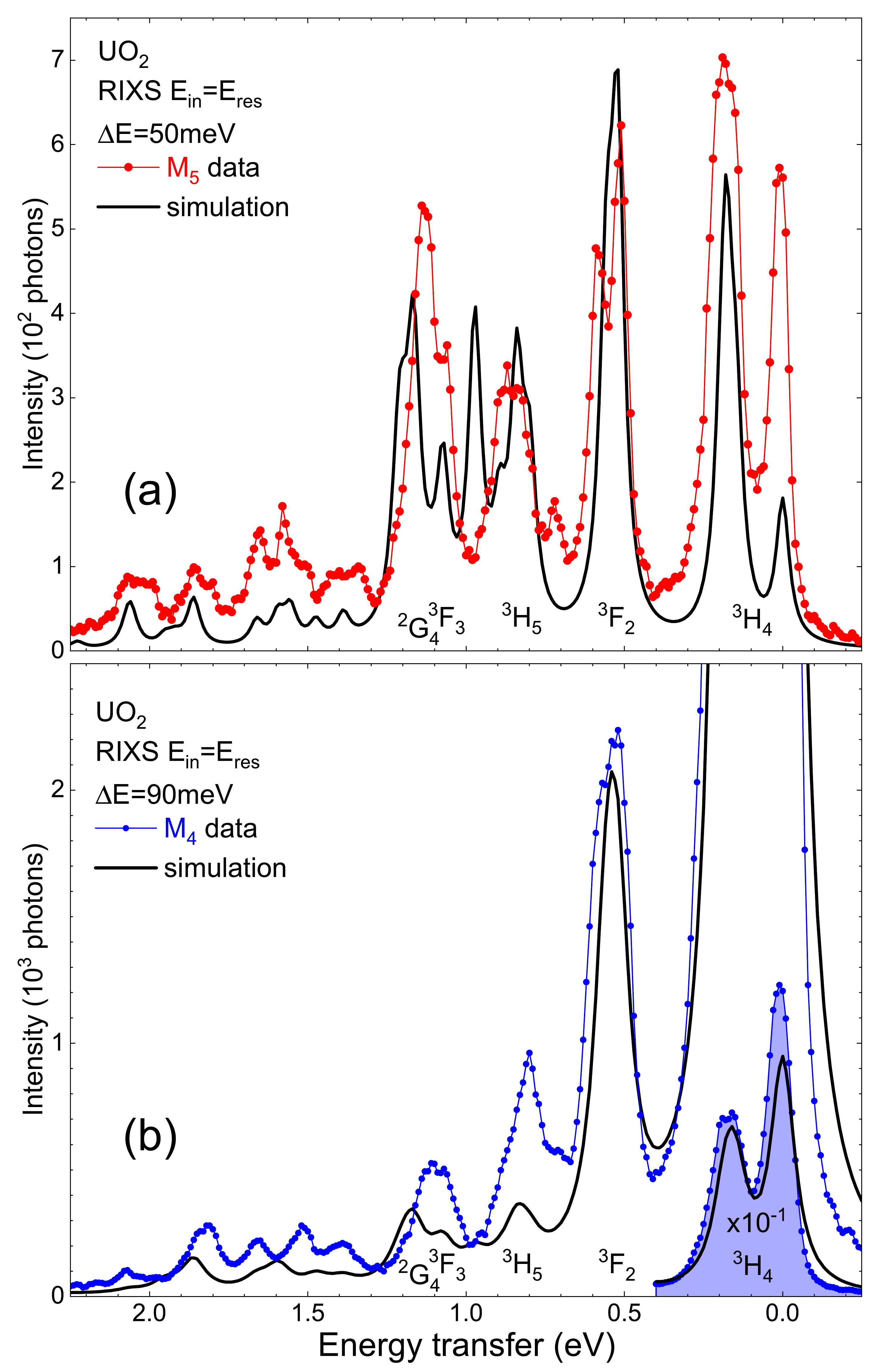}
	\end{center}
	\caption{U\,\textit{M}$_5$-edge (red) data in panel (a) and U\,\textit{M}$_4$-edge (blue) data in panel (b) are the same as in Fig.\,2, shown with expanded intensity scale for the energy transfer range -0.25 to 2.25eV. The inset in the U\,\textit{M}$_4$-edge panel shows the region due to $^3$H$_4$ multiplet scattering divided by 10. LS term symbols are given for orientation. The gray lines correspond to a crystal-field calculation based on a point-charge model (see Sec. III).}  
	\label{RIXS_zoom}
\end{figure}

\begin{figure}[]
	\begin{center}
		\includegraphics[width=0.9\columnwidth]{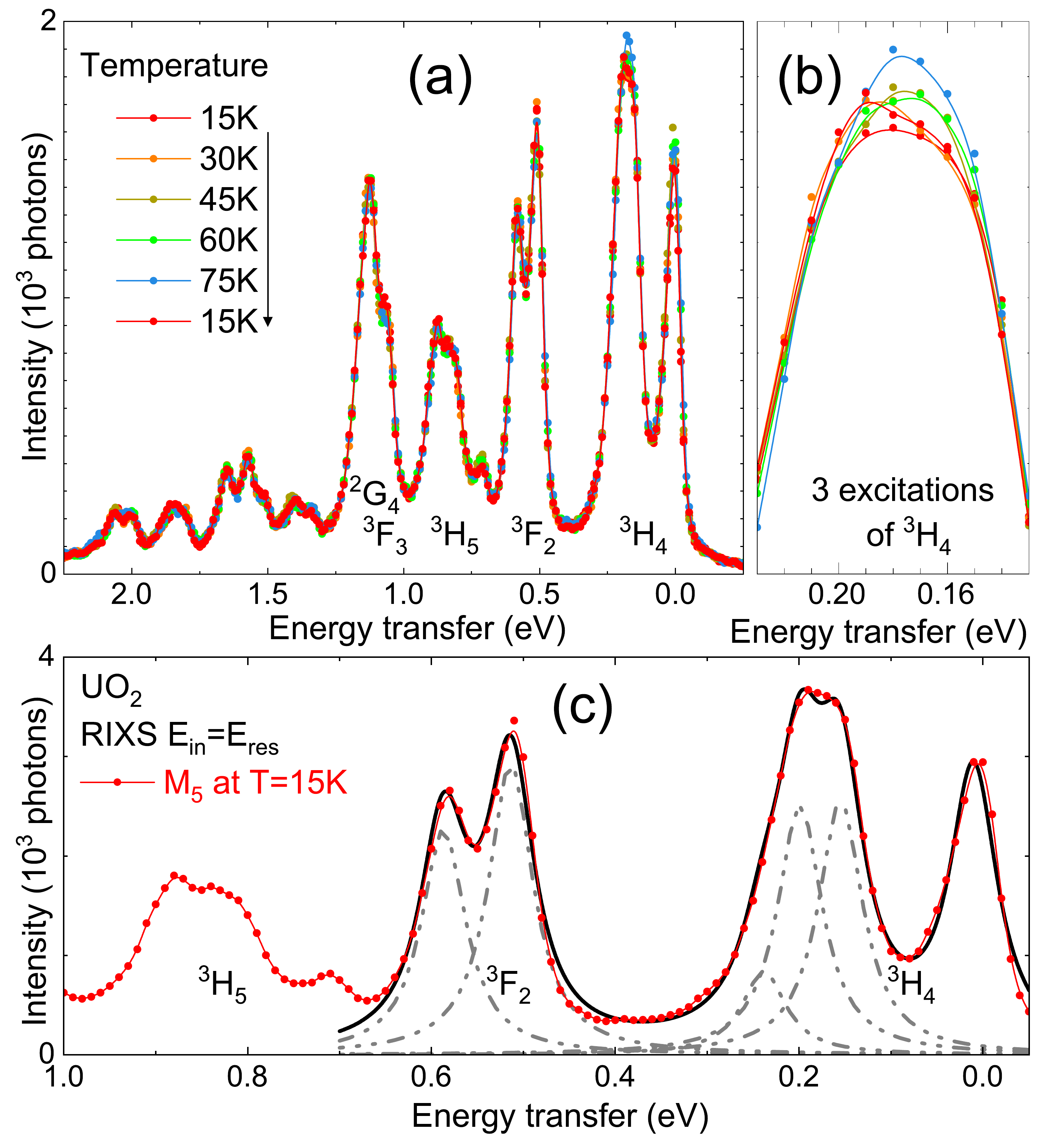}
	\end{center}
	\caption{(a) Temperature dependence of \textit{M}$_5$-edge VB-RIXS data with \textit{E}$_\text{in}$\,=\,\textit{E}$_\text{res}$, measured below and above $T_N$\,=\,30.8\,K, for energy transfer range [-0.25;2.25]eV. The arrow shows the time sequence of the measurements. (b) Zoom to the three crystal-field excitations of the ground state multiplet $^3$H$_4$ in the energy range [0.13;0.23]eV. (c) Empirical Lorentzian fit with equal widths (FWHM\,0.58\,meV) to crystal-field split $^3$F$_2$ (2 lines) and $^3$H$_4$ multiples (3 lines) of \textit{M}$_5$-edge VB-RIXS data at 15\,K. }  
	\label{emp_temp}
\end{figure}

\section{Results and Discussion}
Figure\,\ref{RIXS_all}\,(c), shows the VB-RIXS signal at the \textit{M}$_5$ (red) and \textit{M}$_4$ edges (blue), measured at their respective resonance energies, \textit{E}$_\text{in}$\,=\,\textit{E}$^{\textit{M}5}_{res}$\,=\,3551\,eV and \textit{E}$_\text{in}$\,=\,\textit{E}$^{\textit{M}4}_{res}$\,=\,3725\,eV. The data are normalized to the number of incoming photons (I$_0$), time and energy interval so that intensities are comparable. The data exhibit, in addition to elastic scattering, sharp excitations up to about 2\,eV energy transfer and some additional broad scattering at even higher energies. 

At first we focus at the low energy part of the data. The sharp excitations are assigned to \textit{ff}-excitations (crystal-field split multiplets). We find that the elastic as well as inelastic signal at the \textit{M}$_4$ edge is significantly stronger than at the \textit{M}$_5$ edge. The full multiplet calculation for the U\,5\textit{f}$^2$ configuration in spherical symmetry (see Section Methods), shown in Fig.\,\ref{RIXS_all}\,(d), reveals that this different behavior arises from the VB-RIXS cross-section, i.e. from selection rules being different at the \textit{M}$_5$ and \textit{M}$_4$ edge\,\cite{Laan1996} rather than the different absorption strengths at the two edges. In fact, the absorption at both edges is of the same order of magnitude, it being slightly weaker at the \textit{M}$_4$-edge, see e.g.\,\cite{Wilhelm2018}. 

Figure\,\ref{RIXS_zoom}\,(a) and (b) display the same spectra as in Fig.\,\ref{RIXS_all}\,(c), but for the smaller energy transfer range from -0.25 to 2.25\,eV and with expanded intensity scales. Note, the \textit{M}$_5$-edge y-scale is expanded by a factor of 10 with respect to the \textit{M}$_4$-edge data. The sharp excitations that are assigned to \textit{ff}-excitations are labeled for orientation with the \textit{LS}-coupling term symbols although the intermediate coupling scheme applies. The multiplet excitations occur at the same energy transfers for both edges because they are local electronic excitations, but the cross-sections exhibit strong energy dependence and the \textit{M}$_4$ edge has lower energy resolution so that some excitations remain unresolved. For example, at the \textit{M}$_4$-edge the crystal-field split multiplet excitations $^2$G$_4$ and $^3$F$_3$ are suppressed relative to the $^3$F$_2$ multiplet, whereas at the \textit{M}$_5$-edge, their intensities are approximately equal. This energy-dependent scattering intensity is characteristic for the 5\textit{f}$^2$ configuration and helps identifying the ground state configurations\,\cite{Lander2021,Bright2023}.
%%%%%%%%%%%%%%%%%%%%%%%%%%%%%%%%%%%%%%%%%%%%%%%

Figure\,\ref{emp_temp} presents the temperature (\textit{T}) dependence of the VB-RIXS data measured at the \textit{M}$_5$ resonance for \textit{T}\,=\,15, 30, 45, 60, and 75\,K, followed by another measurement at \textit{T}\,=\,15\,K. Panel (a) shows the energy range from -0.25 to 2.25 and panel (b) a zoomed-in view of the three crystal-field excitations of the ground state multiplet around 0.2\,eV. Only in this energy region a slight temperature dependence is observed upon exiting the antiferromagnetic phase above 30.8\,K, reflecting the intensity redistribution associated with the splitting of the cubic crystal-field levels in the distorted low-temperature phase. The recovery of the spectral shape at 15\,K confirms that the spectral changes are a genuine temperature effect. The incomplete amplitude recovery may stem from a mismatch between sample and nominal temperature. 

\begin{table}
	\caption{Line position (eV) resulting from an empirical fit with Lorentzian lines of equal width 
	}
	\begin{tabular}{l|ccc|cc}
		\hline \hline		
		multiplet     &          &$^3$H$_4$  &           &$^3$F$_2$ &\\
		empirical fit & 0.155 (2)& 0.199 (2) & 0.238 (5) & 0.513 (1) & 0.587 (1) \\ 
		\hline	
	\end{tabular}
\end{table}

In panel (c) of Fig.\,\ref{emp_temp}, we present the result of an empirical fit to the resonant \textit{M}$_5$-edge data at 15\,K using Lorentzian lines of equal width (FWHM\,=\,58\,meV), which is slightly broader than the experimental resolution as determined from the carbon tape. The number of lines corresponds to the number of expected excitations -- three for $^3$H$_4$, which is nine-fold degenerate and split by cubic crystal field into a $\Gamma$$_5$ triplet ground state, along with one excited singlet, doublet and triplet state, and two for  $^3$F$_2$, which is five-fold degenerate and splits into one doublet and one triplet state. The resulting peak positions are listed in Table\,I.  

Inelastic neutron scattering (INS) data with 3\,meV resolution resolves the crystal-field splittings of the $^3$H$_4$ multiplet, revealing substructures induced by magnetic order\cite{Amoretti1989}. The overall agreement between INS and the present \textit{M}-edge VB-RIXS data for the $^3$H$_4$ excitations is good, although we assign peaks at slightly larger energy transfers. The difference may lie in the different selection rules governing RIXS and INS spectra. In the case of INS, dipolar selection rules apply ($\Delta J_z$ = 0, $\pm$1), whereas RIXS permits transitions with ($\Delta J_z$ = 0, $\pm$1, $\pm$2)\,\cite{Moretti2011,Amorese2018}, due to two consecutive dipole transitions. In the ordered phase of UO$_2$, the $\Gamma_5$ triplet ground state is split into one doublet and one singlet state, with the lowest level being the singlet\,\cite{Carretta2010}. Using INS, only three transitions have non-negligible probabilities. For example, the transition between the lowest and highest energy singlets has an intensity too weak to be detected. This limitation does not apply to RIXS due to its wider selection rules. We also would like to note that the higher-energy part of the multiplet structure remain inaccessible in INS due to the weak neutron cross-section.

The $^3$H$_4$ and $^3$F$_2$ multiplets are also resolved in \textit{N}-edge VB-RIXS data, though with a much lower signal to background ratio, despite the better energy resolution of $\Delta$\textit{E}\,=\,30\,meV\,\cite{Lander2021}. The energy positions of the excitations are in good agreement with the present data. In contrast, earlier low-resolution VB-RIXS data at the U\,M$_5$ edge, i.e. with  $\Delta$\textit{E}\,=\,1\,eV, did not resolve the multiplet structure, see Fig.\,12 in Ref.\,\cite{Kvashnina2014}. 

%%%%%%%%%%%%%%%%%%%%crystal-field point-charge simulation%%%%%%%%%%%%%%%%%%%%%%%%%%%%%%%%%%%%%%%%%%%%%%%%%%%%%%%%%%%%%%%%%

Having established in Fig.\,\ref{emp_temp} that temperature effects in the RIXS spectra are negligible, we proceed with a crystal-field simulation that neglects the impact of magnetic order. The gray lines in Figure\,\ref{RIXS_zoom}\,(a) and (b) show the result of such a fit (see description in Section III). At first glance, the agreement appears reasonable. However, a closer look reveals that the simulated crystal-field splittings of $^3$H$_4$ and $^3$F$_2$ multiplets are too small, and the agreement deteriorates further at even higher energy transfers. This cannot be corrected by adjusting the crystal-field parameters V$_{40}$ and V$_{60}$, or varying the Slater integrals or spin-orbit coupling parameters (see Section III), because the observation of several excited multiplets puts very strong restrictions to the Slater integrals and spin-orbit coupling in the single-ion model. Instead, the discrepancies suggest that the point-charge model is too simplistic and that the effect of hybridization with the surrounding ligands (which also contributes to the crystal-field effect) must be taken into account explicitly. This consideration brings us to the high energy part of the RIXS spectrum, where extra intensities are observed.

%%%%%%%%%%%%%%%%%%%High energy part  -- charge transfer %%%%%%%%%%%%%%%%%%%%%%%%%%%%

\begin{figure}[]
	\begin{center}
		\includegraphics[width=0.99\columnwidth]{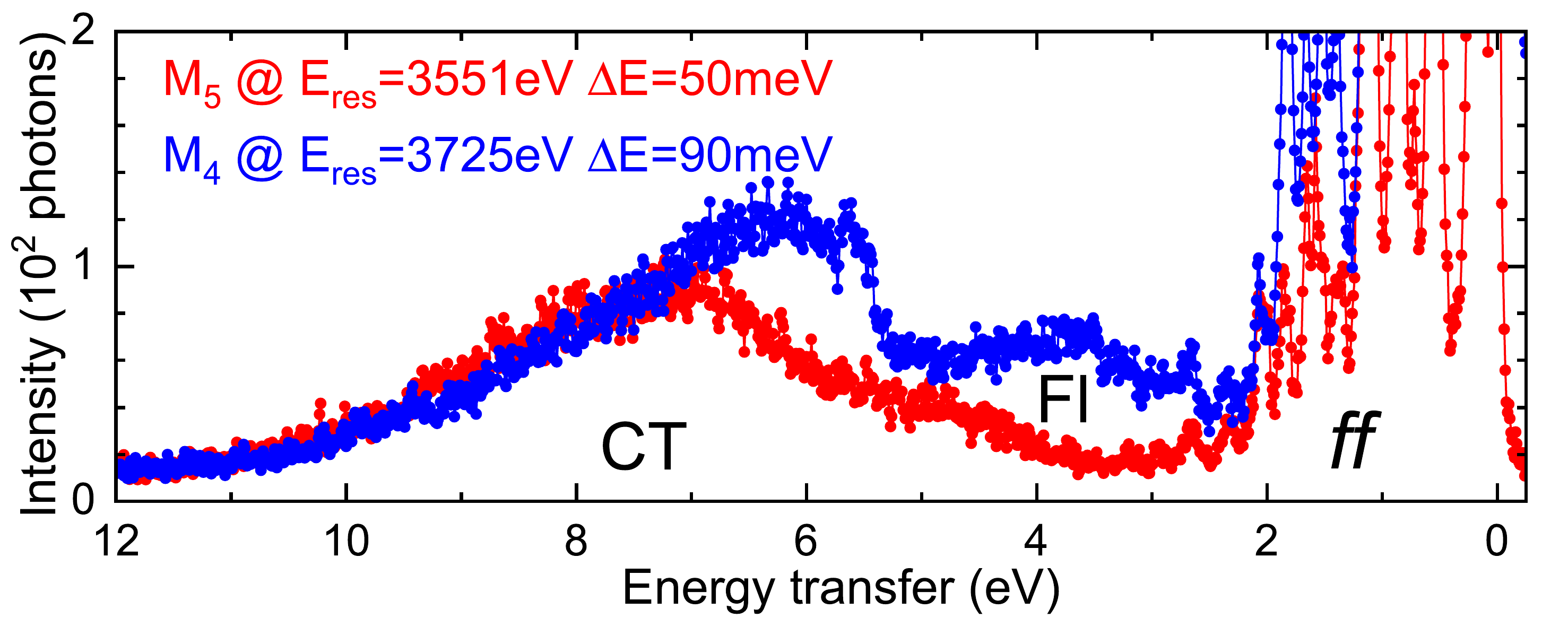}
	\end{center}
	\caption{Intensity zoom to high energy excitations of VB RIXS data measured at \textit{M}$_5$ (red) and \textit{M}$_4$ edge (blue) with \textit{E}$_\text{in}$\,=\,\textit{E}$_\text{res}$. Multiplet exciatitons are denoted with \textbf{\textit{ff}}, fluorescence with \textbf{Fl} and charge transfer with \textbf{CT}. }  
	\label{long}
\end{figure}

\begin{figure}[]
	\begin{center}
		\includegraphics[width=0.99\columnwidth]{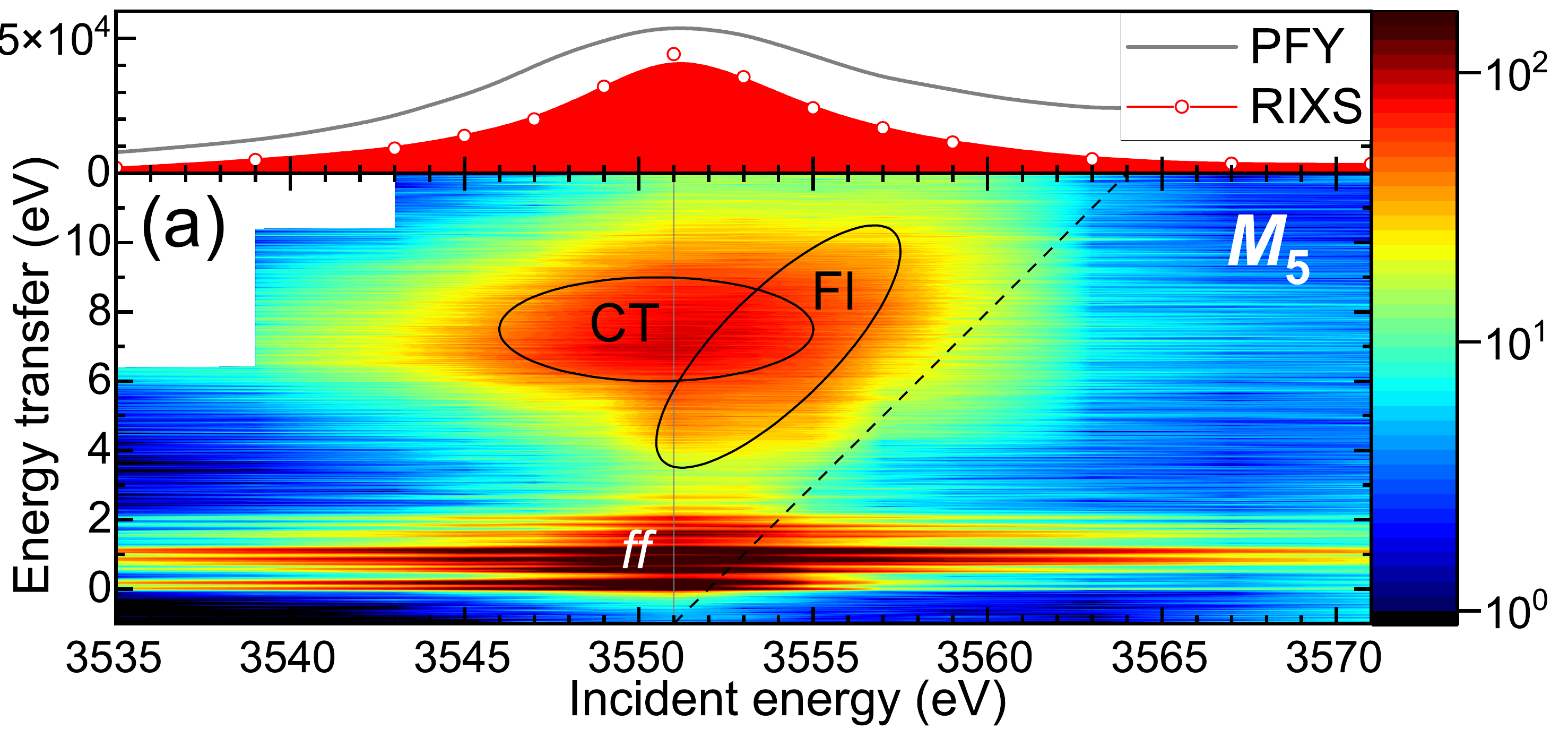}
		\includegraphics[width=0.99\columnwidth]{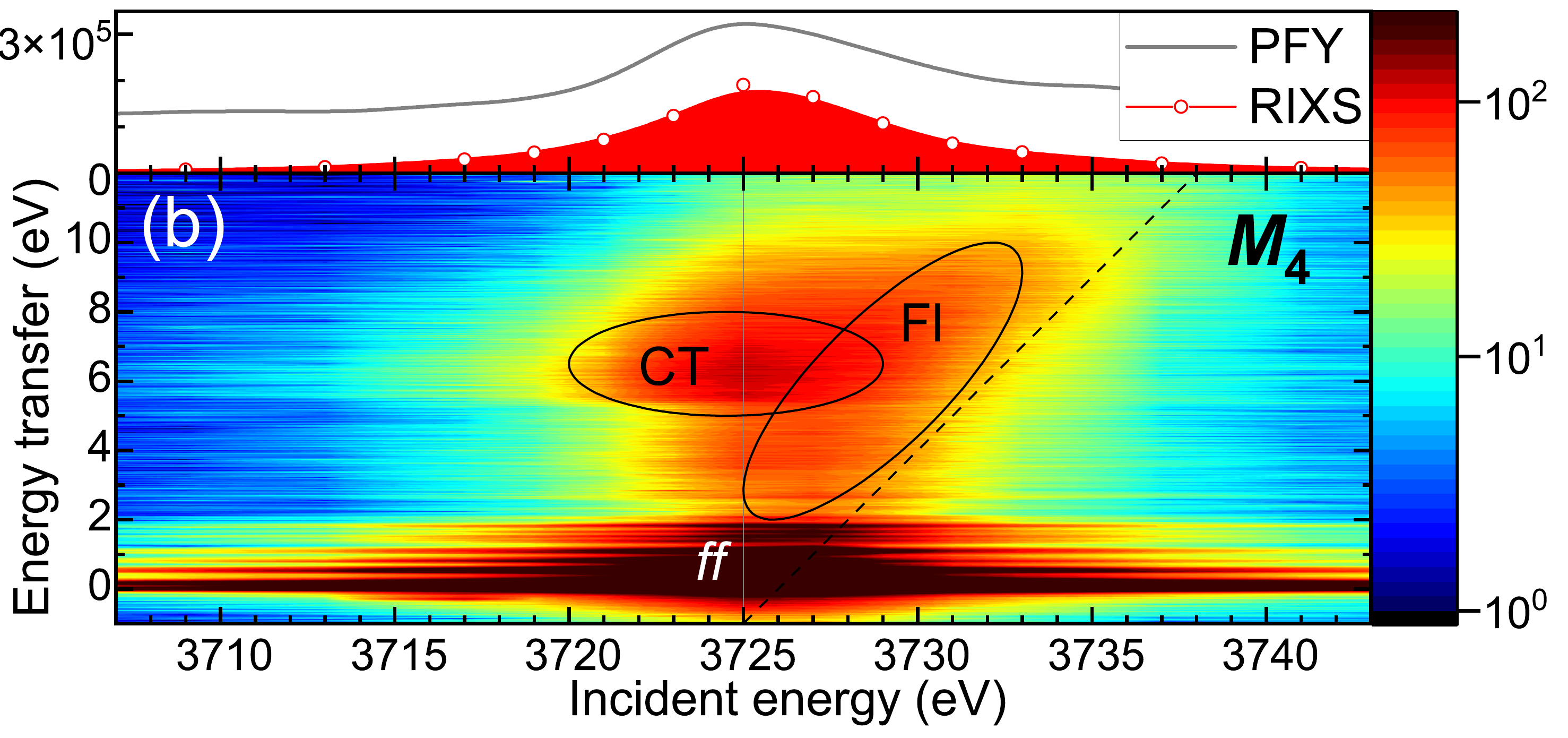}
	\end{center}
	\caption{Energy transfer -- incident energy maps for the \textit{M}$_5$- (a) and \textit{M}$_4$-edges (b). On top the integrated intensity over all energy transfers (red) compared to XAS-data (grey). The dashed line refers to constant final energy starting at \textit{E}$_{res}$, respectively. The ellipses mark regions of charge transfer (\textbf{CT}) and fluorescence (\textbf{Fl}), and \textit{\textbf{ff}} marks regions of multiplet scattering.}  
	\label{map}
\end{figure}

Before turning to the high energy part of the RIXS spectrum, we note that, when utilizing resonances, two types of excitation processes can be distinguished, Raman-like and fluorescence-like. Multiplet or charge-transfer excitations exhibit Raman-like behavior: they appear at a constant energy transfer as the incident energy \textit{E}$_{in}$ is scanned across the resonance. In contrast, fluorescence-like excitations are characterized by a fixed final energy \textit{E}$_{out}$, so that the energy transfer varies with \textit{E}$_{in}$\cite{Kotani2001,Ament2011}. This distinction is analogous to the participator and spectator processes observed in resonant photoemission, where spectral features can be attributed to either direct photoemission (participator) or Auger-like decay (spectator)\,\cite{Tjeng1991, Tjeng1993}.

The high energy part of the VB-RIXS data is shown in Fig.\,\ref{long} for the \textit{M}$_5$ and \textit{M}$_4$ edges at resonance, respectively, on a blown up intensity scale. We observe strong scattering of about equal strength at both edges, despite the significant difference in the RIXS cross-sections for the multiplet scattering (compare with Fig.\,\ref{RIXS_all}). For both edges, this additional scattering is very broad, and exhibits two features at about 4.5 and 7\,eV in the \textit{M}$_5$- and about 3.75 and 6.25\,eV in the \textit{M}$_4$-edge data, respectively. This high energy scattering is not accounted for by the full multiplet calculations as shown in Fig.\,\ref{RIXS_all}\,(c), and must be assigned to charge transfer or fluorescence excitations. 

A more systematic view on the overall VB RIXS spectra is presented in the energy transfer vs. incident energy maps in Fig.\,\ref{map}\,(a) and (b). These maps were generated from spectra taken for  different incident energies across the respective absorption edges, covering a range of more than 35\,eV, see Fig.\,\ref{RIXS_all}\,(c) and (d). Additionally, we display the XAS or, to be more exact, the partial fluorescence yield (PFY) signal at the top (grey line), compared to the RIXS signal that corresponds to the integral over all energy transfers (red signal). 

The maps show that the scattering intensity associated with local (\textit{ff}) multiplet excitations, up to about 2\,eV energy transfer, exhibit Raman-like behavior, i.e., the energy transfer does not vary with incident energy. However, scattering intensities at higher energy transfers, from 3-4\,eV up to 10\,eV, show an asymmetric intensity distribution. This higher energy region can be decomposed into two components:\\ (1) A Raman-like contribution, peaking at energy transfers of 7\,eV (\textit{M}$_5$) and 6.25\,eV (\textit{M}$_4$) at their respective resonance energies. These features, labeled CT in Fig.,\ref{map}(a) and (b), are attributed to charge-transfer excitations.\\ (2) A fluorescence-like contribution that appears at and above the respective resonance energies and shifts to higher energy transfer with increasing incident energy. These features are labeled Fl.

Interestingly, the fluorescence-like signal is weaker at the \textit{M}$_5$ edge compared to the \textit{M}$_4$ edge. This disparity may be linked to the higher photon energy of the \textit{M}$_4$ edge, which potentially opens more decay channels. Similar behavior has been observed in \textit{L}$_{2,3}$ valence-band RIXS measurements of transition metal oxides \cite{Ghiringhelli2009, Bisogni2016, Hariki2018, Winder2020, Hariki2020}.

Crucial for the detection of charge-transfer excitations is the identification of a window of incident energies where the VB-RIXS signal remains predominantly Raman-like, with minimal or no fluorescence-like contributions. This condition is fulfilled by tuning the incident energy below the resonance and by using the \textit{M}$_5$ edge rather than the \textit{M}$_4$. For measuring and later simulating the multiplets, using both the \textit{M}$_5$ and \textit{M}$_4$ edges is advantageous, as it allows full exploitation of the selection rules governing the VB-RIXS process.

The description of the entire spectrum, low as well as high energy, in one go should be the ultimate goal to obtain a comprehensive description of the U\,5$f$ electronic structure, i.e., the degree of covalency as well as crystal-field level scheme. This requires incorporating both multiplet effects and covalency, including the presence of multiple U\,5$f$ configurations, which is an ambitious challenge for theory. However, a key advantage  of VB-RIXS is that the parameters in the initial (ground) and final state are identical, as illustrated in Fig.\,\ref{method}\,(a)\&(c).  The present data set also offers an excellent opportunity to test and refine more recent theories of UO$_2$, see Ref.\,\cite{Butorin2016,Ramanantoanina2016,Koloren2018} among others, developed for the interpretation of core-to-core RIXS data, where multiplet structures are typically unresolved. More generally, we believe that the more detailed information accessible through high-resolution \textit{M}$_{4,5}$-edge experiments has the potential for advancing the understanding of actinide materials.

It should be emphasized that \textit{M}-edge VB-RIXS is not limited to semiconducting actinide materials. It can also be applied to intermetallic uranium compounds, where excitations are typically not observed in inelastic neutron scattering. References \cite{Marino2023,Christovam2024,Marino2024,SundermannUTe2} have demonstrated that multiplet structures appear in narrow 5\textit{f}-band intermetallic compounds. In contrast, such features are absent in band-like materials where the 5\textit{f} shells overlap. Between these limits, the dual character of the 5\textit{f} electrons, localization versus itinerancy, governs the fascinating properties of many actinide compounds. In VB-RIXS, this duality manifests as the competing coexistence of atomic-like 5\textit{f} multiplet structures and a 5\textit{f}-band response, the latter originating from direct 5\textit{f}–5\textit{f} overlap and/or indirect hybridization with ligand states.

\section{Summary} 
The present \textit{M}-edge VB-RIXS study on UO$_2$ shows that electronic excitations in uranium compounds, ranging  from low to high energies, can be resolved with unprecedented detail. These include crystal-field-split multiplet excitations up to approximately 2\,eV, as well as charge-transfer excitations in the 3 to 10\,eV range. All these excitations are detected without a core-hole in the final state. The successful simulation of the spectra will provide reliable information about the symmetry, the crystal-field level scheme, and the degree of covalency or intermediate/mixed-valent character of the U\,5\textit{f} electrons in the initial (ground) state. These insights are made possible by the excellent signal-to-background ratio at the U\,\textit{M}$_{4,5}$-edges and the state-of-the-art resolution of 50\,meV. \textit{M}-edge VB-RIXS therefore emerges as a powerful and promising tool for probing the electronic structure of uranium and, more broadly, actinide compounds.

\section{Acknowledgment}
All authors thank Philippe Raison from the Joint Research Centre of the European Comission in Karlsruhe, Germany for providing the UO$_2$ single crystal and acknowledge DESY (Hamburg, Germany), a member of the Helmholtz Association HGF, for the provision of experimental facilities. A.S. is thankful for insightful discussions with Lucia Amidani and is grateful for support from the German Research Foundation (DFG) - grant N$^{\circ}$ 387555779. B.K. acknowledges funding from the European Research Council (ERC) under Advanced Grant No. 331 101141844 (SpecTera).

%\bibliography{references.bib} 

\end{document}